\documentstyle[12pt,epsf,axodraw]{article}
\begin{document}
\phantom{.}
\vskip 2truemm
\begin{center}
\Large
{\bf Check of the Bootstrap Conditions \\
for the Gluon Reggeization}
\end{center}
\vskip 0.3cm
\centerline{Alessandro Papa}
\vskip .3cm
\centerline{\sl Dipartimento di Fisica, Universit\`a della Calabria
\& INFN - Cosenza}
\centerline{\sl I-87036 Arcavacata di Rende (Cosenza), Italy}
\vskip .5cm
\begin{abstract}
The property of gluon Reggeization plays an essential role in the derivation 
of the Balitsky-Fadin-Kuraev-Lipatov (BFKL) equation for the cross sections at 
high energy $\sqrt s$ in perturbative QCD. This property has been proved to 
all orders of perturbation theory in the leading logarithmic approximation and
it is assumed to be valid also in the next-to-leading logarithmic approximation,
where it has been checked only to the first three orders of perturbation theory.
From $s$-channel unitarity, however, very stringent ``bootstrap'' conditions 
can be derived which, if fulfilled, leave no doubts that gluon Reggeization holds.
\end{abstract}

\vskip.3cm

The BFKL equation~\cite{Fad00} is very important for the theory of 
Regge processes at high energy $\sqrt s$ in perturbative QCD,
such as the deep inelastic $e\,p$ scattering in the region of 
small values of the Bjorken variable $x$, presently investigated 
at HERA.
It was derived more than twenty years ago in the leading logarithmic 
approximation (LLA)~\cite{Fad00}, which means summation of all the terms of the 
type $(\alpha_s \ln s)^n$. Recently, the radiative corrections to the equation 
have been calculated (see Ref.~\cite{Fad00} for a review) and the explicit form of 
the kernel of the equation in the next-to-leading approximation (NLA) for the 
case of forward scattering became known~\cite{Fad00}. 

The key role in the derivation of the BFKL equation is played by the gluon 
Reggeization. ``Reggeization'' of a given elementary particle usually means that
the amplitudes of the scattering processes with exchange of the quantum numbers 
of that particle in the $t$-channel go like $s^{j(t)}$ in the 
Regge limit $s\gg |t|$. The function $j(t)$ is called ``Regge trajectory''
of the given particle and takes the value of the spin of that particle
when $t$ is equal to its squared mass. It is well-known that in QED the electron
Reggeizes in perturbation theory~\cite{GGLMZ}, while the photon remains 
elementary~\cite{Man64}, while in QCD the gluon Reggeizes~\cite{GS73,Lip76} 
as well as the quark~\cite{FS77}. In perturbative QCD, the notion of gluon 
Reggeization is used in a stronger sense. It means not only that a Reggeon 
exists with the quantum numbers of the gluon, negative signature and with a 
trajectory $j(t) = 1 + \omega(t)$ passing through 1 at $t=0$, 
but also that this Reggeon gives the leading contribution in each order of 
perturbation theory to the amplitudes of the processes with large $s$ 
and fixed (i.e. not growing with $s$) squared momentum transfer $t$.

The simplest realization of the gluon Reggeization is in the elastic process 
$ A + B \longrightarrow A^\prime + B^\prime $ with exchange of gluon quantum 
numbers in the $t$-channel (see Fig.~1), whose amplitude in the Regge limit 
takes the form
\begin{equation}
\left({{\cal A}_8^-}\right)^{A^\prime B^\prime}_{AB} = 
\Gamma^c_{A^\prime A}\:\left[\left({-s\over -t}\right)^{j(t)}-\left({s\over 
-t}\right)^{j(t)}\right]\:\Gamma^c_{B^\prime B}\;.
\label{elast_ampl_8}
\end{equation}
Here $c$ is a color index and $\Gamma^c_{P^\prime P}$ are the 
particle-particle-Reggeon (PPR) vertices, not depending on $s$. They can be
written in the form $\Gamma^{c}_{P^\prime P} = g\langle P^\prime| T^c| P \rangle 
\Gamma_{P^\prime P}$, where $g$ is the QCD coupling constant and $T^c$ are the 
color group generators in the fundamental (adjoint) representation for quarks 
(gluons)\footnote{Throughout this paper only the case of quarks and gluons
as colliding particles is considered.}. In the LLA this form of the amplitude 
has been rigorously 
proved~\cite{BFL79}: $\Gamma^{(0)}_{P^\prime P}$ is given simply by
$\delta_{{\lambda_{P^\prime}}{\lambda_P}}$, where
$\lambda_P$ is the helicity of the particle $P$, and the Reggeized gluon
trajectory is needed with 1-loop accuracy~\cite{Fad00},
\begin{equation}
\omega^{(1)}(t) = {g^2 t\over {(2{\pi})}^{D-1}}\frac{N}{2}
\int{d^{D-2}k_\perp\over k_\perp^2{(q-k)}_{\perp}^2}\;.
\label{omega_1}
\end{equation}
Here $D=4+2 \epsilon$ has been introduced in order to regularize the infrared 
divergences and the integration is performed in the space transverse to the 
momenta of the initial colliding particles. In the NLA (resummation of 
the terms $\alpha_s^{n+1} (\ln s)^n$), the form (\ref{elast_ampl_8}) has been
checked in the first three orders of perturbation theory~\cite{Fad00} and is
only assumed to be valid to all orders. In this approximation, the NLA
contribution to the PPR vertices is needed, which takes the form
$\Gamma^{(1)}_{P^\prime P} = 
\delta_{\lambda_{P^\prime}\lambda_P}\Gamma^{(+)}_{PP}+
\delta_{\lambda_{P^\prime},-\lambda_P}\Gamma^{(-)}_{PP}$, where a 
helicity non-conserving term appeared, and the Reggeized gluon trajectory enters
with 2-loop accuracy. 

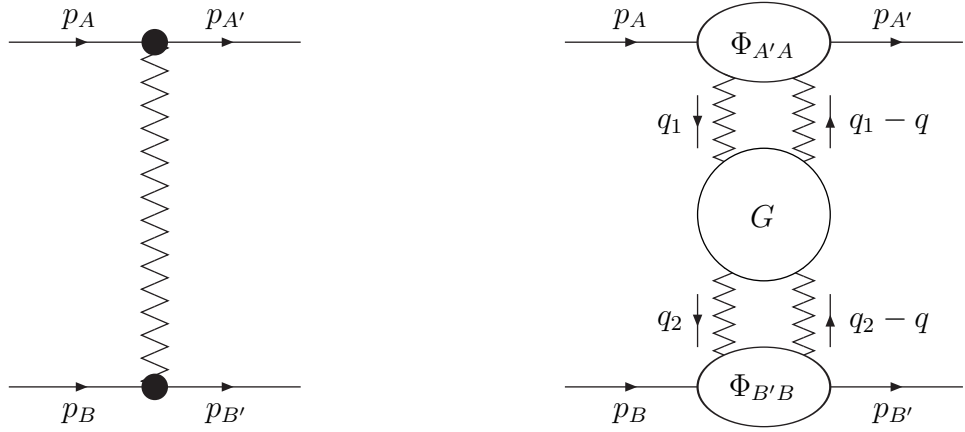
\begin{figure}[tb]
\vspace{3truemm}
\begin{minipage}{80mm}
\begin{center}
\setlength{\unitlength}{0.35mm}
\begin{picture}(150,150)(0,0)
\ArrowLine(20,140)(75,140)
\ArrowLine(75,140)(130,140)
\Vertex(75,140){5}
\ArrowLine(20,10)(75,10)
\ArrowLine(75,10)(130,10)
\Vertex(75,10){5}
\ZigZag(75,140)(75,10){5}{15}
\Text(47,150)[c]{$p_A$}
\Text(47,0)[c]{$p_B$}
\Text(103,150)[c]{$p_{A^\prime}$}
\Text(103,0)[c]{$p_{B^\prime}$}
\end{picture}
\end{center}
\end{minipage}
\begin{minipage}{80mm}
\begin{center}
\setlength{\unitlength}{0.35mm}
\begin{picture}(150,150)(0,0)
\ArrowLine(0,140)(50,140)
\ArrowLine(100,140)(150,140)
\Text(25,150)[c]{$p_A$}
\Text(125,150)[c]{$p_{A'}$}
\Oval(75,140)(15,25)(0)
\Text(75,140)[c]{$\Phi_{A'A}$}
\ZigZag(60,128)(60,95){4}{5}
\ZigZag(90,128)(90,95){-4}{5}
\ZigZag(60,22)(60,55){4}{5}
\ZigZag(90,22)(90,55){-4}{5}
\ArrowLine(50,120)(50,100)
\ArrowLine(50,45)(50,25)
\ArrowLine(100,100)(100,120)
\ArrowLine(100,25)(100,45)
\Text(45,110)[r]{$q_1$}
\Text(108,111)[l]{$q_1-q$}
\Text(45,35)[r]{$q_2$}
\Text(108,35)[l]{$q_2-q$}
\GCirc(75,75){25}{1}
\Text(75,75)[]{$G$}
\ArrowLine(0,10)(50,10)
\ArrowLine(100,10)(150,10)
\Text(25,0)[]{$p_B$}
\Text(125,0)[]{$p_{B'}$}
\Oval(75,10)(15,25)(0)
\Text(75,10)[c]{$\Phi_{B'B}$}
\end{picture}
\end{center}
\end{minipage}
\vspace{0.2cm}
\caption[]{(Left) Diagrammatical representation of 
$({{\cal A}_8^-})^{A^\prime B^\prime}_{AB}$. The zig-zag line
is the Reggeized gluon, the black blobs the PPR effective vertices. 
(Right) Diagrammatical representation of 
$({\cal A}_{\cal R})_{AB}^{A^{\prime }B^{\prime }}$ from $s$-channel unitarity. 
The ovals are the impact factors of the particles $A$ and $B$, the circle is 
the Green function for the Reggeon-Reggeon scattering.}
\end{figure}

On the other hand, the amplitude for the elastic scattering process 
$ A + B \longrightarrow A^\prime + B^\prime $ (for arbitrary color representation
in the $t$-channel) can be determined from $s$-channel unitarity, by 
expressing its imaginary part in terms of the inelastic amplitudes
$A + B \longrightarrow \tilde A + \tilde B + \{n\}$ and 
$A^\prime + B^\prime \longrightarrow \tilde A + \tilde B + \{n\}$, and then
by reconstructing the full amplitude by use of the dispersion relations. 
Of course, the additional particles $\{n\}$ to be considered in the intermediate
states and their kinematical configurations depend on the adopted approximation. 
It turns out (see for instance~\cite{Fad00,FF98}) that this amplitude can 
be written as (see Fig.~1)
\begin{equation}
({\cal A}_{\cal R})_{AB}^{A^{\prime }B^{\prime }}=
\frac{i \, s}{\left( 2\pi \right) ^{D-1}}\int \frac{d^{D-2}q_{1}}{\vec{q}
_{1}^{\:2} \vec{q}_{1}^{\:\prime\:2}}\int \frac{d^{D-2}q_{2}}{\vec{
q}_{2}^{\:2} \vec{q}_{2}^{\:\prime\: 2}}\int_{\delta-i\infty}^{\delta+i\infty} 
\frac{d\omega}{\mbox{sin}(\pi\omega)}
\left[ \left( \frac{-s}{s_{0}}\right)^{\omega}
-\tau\left( \frac{s}{s_{0}}\right)^{\omega}\right]
\label{elast_ampl_R}
\end{equation}
\[
\times \sum_{ {\cal R},\nu} \Phi _{A^{\prime
}A}^{\left( {\cal R},\nu \right) }\left( \vec{q}_{1},\vec{q};s_{0}\right)
G_{\omega }^{\left( {\cal R}\right) }\left( \vec{q}_{1},\vec{q}_{2},\vec{q}\right)
\Phi _{B^{\prime }B}^{\left( {\cal R},\nu \right) }
\left( -\vec{q}_{2},-\vec{q};s_{0}\right)\,,
\]
where ${\cal A}_{\cal R}$ stands for the scattering amplitude with
the representation ${\cal R}$ of the color group in the $t$-channel, 
the index $\nu$ enumerates the states in the irreducible representation 
${\cal R}$ and $G_{\omega }^{\left( {\cal R}\right)}$ is the Mellin 
transform of the Green function for the Reggeon-Reggeon scattering~\cite{FF98}.
The signature $\tau$ is positive (negative) for symmetric (antisymmetric) 
representation ${\cal R}$. The parameter $s_0$ is an arbitrary energy scale 
introduced in order to define the partial wave expansion of the scattering 
amplitudes through the Mellin transform. The dependence on this parameter 
disappears in the full expressions for the amplitudes. 
$\Phi_{P^\prime P}^{\left( {\cal R},\nu \right)}$ are the so-called impact factors, 
defined by
\begin{equation}
\Phi _{P^{^{\prime }}P}^{\left( {\cal R}\mbox{,}\nu \right) }\left( \vec{q}
_{R}\mbox{,}\vec{q}\mbox{;}s_{0}\right) =\int \frac{ds_{PR}}{2\pi s}
{\mathrm Im}{\cal A}_{P^{^{\prime }}P}^{\left( {\cal R},\nu \right) }
\left(p_{P},q_{R};\vec{q};s_{0}\right) \theta 
\left( s_\Lambda-s_{PR}\right) 
\label{imp_fact}
\end{equation}
\[
-\frac{1}{2}\int \frac{d^{D-2}q^{\prime}}{\vec{q}^{\:\prime 2}\left( \vec{q}
^{\:\prime }-\vec{q}\right) ^{2}}\Phi _{P^{^{\prime }}P}^{\left( {\cal R},\nu 
\right) }\left( \vec{q}^{\:\prime },\vec{q}\right) 
{\cal K}_{r}^{\left( {\cal R} \right) B }
\left( \vec{q}^{\:\prime },\vec{q}
_{R}\right) \mbox{ln}\left( \frac{s_\Lambda^{2}}{\left( \vec{q}
^{\:\prime }-\vec{q}_{R}\right) s_{0}}\right)\;,
\]
where ${\cal A}_{P^{^{\prime }}P}$ is the particle-Reggeon scattering amplitude.
This definition is valid both in the LLA and in the NLA. In the former case, 
the second term in the R.H.S. of the above equation as well as
the $\theta$ function in the first term are not active. In the NLA, instead, 
the second term
behaves as a counterterm for the large $s_{PR}$ contribution to the first
integral, already taken into account in the LLA. The parameter $s_\Lambda$ 
disappears in the final expression for the NLA impact factors.

The Green function obeys the generalized BFKL equation 
\begin{equation}
\omega G_{\omega }^{\left( {\cal R}\right) }\left( \vec{q}_{1},\vec{q}_{2},
\vec{q}\right) = \vec{q}_{1}^{\:2}\vec{q}_{1}^{\:\prime\:2}
\delta ^{\left(
D-2\right) }\left( \vec{q}_{1}-\vec{q}_2\right)
 +\int \frac{d^{D-2}{q}_r}{\vec{q}_r^{\:\prime\:2}\vec{q}_r^{\:2}}
{\cal K}^{\left( {\cal R}\right) }\left( \vec{q}_{1},\vec{q}_r;\vec{q}
\right) G_{\omega }^{\left( {\cal R}\right) }\left( \vec{q}_r,\vec{q}
_{2};\vec{q}\right)\;,
\end{equation}
where we have introduced the notation $q_i^\prime \equiv q_i - q$. 
The kernel ${\cal K}^{( {\cal R})}$ consists 
of two parts, a ``virtual'' part, related with the Reggeized gluon trajectory, and 
a ``real'' part, related to particle production:
\begin{equation}
{\cal K}^{\left( {\cal R}\right) }\left( \vec{q}_{1},\vec{q}_{2};\vec{q}
\right) =\left[ \omega \left( -\vec{q}_{1}^{\:2}\right) +\omega \left( -
\vec{q}_{1}^{\:\prime\:2}\right) \right] \: \vec{q}_{1}^{\:2} \:
\vec{q}_{1}^{\:\prime\:2}\: \delta ^{\left( D-2\right)
}\left( \vec{q}_{1}-\vec{q}_{2}\right) +{\cal K}_{r}^{\left( {\cal R}\right)
}\left( \vec{q}_{1},\vec{q}_{2};\vec{q}\right)  
\end{equation}
The ``real'' part of the kernel is defined by
\begin{equation}
{\cal K}_{r}^{\left( {\cal R}\right) }\left( \vec{q}_{1},\vec{q}_{2};\vec{q}
\right) =\int \frac{ds_{RR}}{\left( 2\pi \right)^{D}}{\mathrm Im}
{\cal A}_{RR}^{\left( {\cal R}\right) }\left( q_{1},q_{2};\vec{q}\right) 
\theta \left( s_\Lambda-s_{RR}\right) 
\label{kernel}
\end{equation}
\[
-\frac{1}{2}\int \frac{d^{D-2}q^{\prime}}{\vec{q}^{\:\prime 2}\left( \vec{q}
^{\:\prime }-\vec{q}\right) ^{2}}{\cal K}_{r}^{\left( {\cal R}\right) B}
\left( \vec{q}_{1},\vec{q}^{\:\prime };\vec{q}\right) 
{\cal K}_{r}^{\left( {\cal R}\right) B}
\left( \vec{q}^{\:\prime },\vec{q}_{2};\vec{q}\right) 
\mbox{ln}\left( \frac{s_\Lambda^{2}}{\left( \vec{q}^{\:\prime }-\vec{q}
_{1}\right) ^{2}\left( \vec{q}^{\:\prime }-\vec{q}_{2}\right) ^{2}}\right)\;,  
\]
where ${\cal A}_{RR}^{\left( {\cal R}\right) }$ is the Reggeon-Reggeon 
scattering amplitude.
Here the same arguments concerning $s_\Lambda$ apply as in the discussion 
after Eq.~(\ref{imp_fact}).
In the LLA, the Reggeized gluon trajectory is needed at 1-loop accuracy and the 
only contribution to the ``real'' part of the kernel is from the production 
of one gluon at Born level in the collision of two Reggeons 
(${\cal K}_{RRG}^{(B)}$).
In the NLA, the Reggeized gluon trajectory is needed at 2-loop accuracy and the 
``real'' part of the kernel takes contributions from one-gluon production at 
1-loop level (${\cal K}_{RRG}^{(1)}$) and from two-gluon and $q \overline 
q$-pair production at Born level (${\cal K}_{RRGG}^{(B)}$ and 
${\cal K}_{RRQ\overline Q}^{(B)}$, respectively). The representation 
(\ref{elast_ampl_R})
for the elastic amplitude must reproduce with NLA accuracy the representation
(\ref{elast_ampl_8}), in the case of exchange of gluon quantum numbers (i.e.
color octet representation and negative signature) in the $t$-channel.
This leads to two very stringent ``bootstrap'' conditions~\cite{FF98}:
\begin{equation}
\frac{g^2Nt}{2\left( 2\pi \right) ^{D-1}}\int \frac{d^{D-2}q_1}{\vec
q_1^{\:2}\vec q_1^{\:\prime\:2}}\int \frac{d^{D-2}q_2}{\vec
q_2^{\:2}\vec q_2^{\:\prime\:2}}{\cal K}^{\left(8\right) \left(1\right)
}\left( \vec q_1,\vec q_2;\vec q\right)
=\omega ^{\left( 1\right) }\left( t\right) \omega ^{\left( 2\right) }
\left( t\right)\;,
\label{boot1}
\end{equation}
\begin{equation}
\frac{ig\sqrt{N}t}{(2\pi)^{D-1}}\int\frac{d^{D-2}q_1}{\vec q_1^{\:2}
\vec q_1^{\:\prime\:2}}\Phi_{A^{\prime}A}^{(8,a)(1)}(\vec q_1, \vec q; 
s_0) \!=\! \Gamma_{A^{\prime}A}^{a(1)}\,\omega^{(1)}(t)
\!+\! \frac{\Gamma_{A^{\prime}A}^{a(B)}}{2} \left[ \omega^{(2)}(t) 
\!+\! \left( \omega^{(1)}(t) \right)^2\ln\left(\frac{s_0} {\vec q^{\:2}} \right) 
\right].
\label{boot2}
\end{equation}
The first of them involves the kernel of the generalized non-forward
BFKL equation in the octet color representation and at 1-loop order,
together with the 1- and 2-loop contributions to the Reggeized gluon trajectory;
the second one involves the impact factors in the octet color representation
at 1-loop order together with the 1- and 2-loop contributions to the Reggeized gluon
trajectory and the PPR effective vertices at Born and 1-loop level. 
Besides providing a stringent check of the gluon Reggeization in the NLA, 
the above equations are important since they test, at least in part, the 
correctness of the year-long calculations which lead to the NLA BFKL equation.

The first bootstrap condition has been considered for the part concerning 
the quark contribution (massless case)~\cite{FFP99}. The ingredients for this
equation are the quark contribution to the 2-loop Reggeized gluon 
trajectory $\omega_Q^{(2)}(t)$~\cite{Fad00}
and the quark contribution to the ``real'' part of the NLA kernel in the octet
color representation, 
${\cal K}_r^{Q \,(8)\,(1)}( \vec q_1,\vec q_2;\vec q)$,
\[
{\cal K}_r^{Q \, (8)\,(1)}\left( \vec q_1,\vec q_2;\vec q\right) = 
{\cal K}_{RRG} ^{Q \,(8)\,(1)}\left( \vec q_1,\vec q_2;
\vec q\right) + {\cal K}_{RRQ\overline{Q}}^{\left( 8\right) \left(B\right) }
\left( \vec q_1,\vec q_2;\vec q\right)\;.
\]
${\cal K}_{RRG} ^{Q \,(8)\,(1)}$ depends on the RRG effective vertex at Born 
level, $\gamma_{c_1 c_2}^{G (B)}$~\cite{Fad00}
and on the real part of the quark contribution to the RRG effective vertex 
at 1-loop level, $\gamma_{c_1 c_2}^{G\,(Q)(1)}$~\cite{Fad00}; 
${\cal K}_{RRQ\overline{Q}}^{\left( 8\right) \left(B\right) }$ depends instead
on the RRQ$\overline Q$ effective vertex, 
$\gamma _{c_1c_2}^{Q{\overline{Q}}}$~\cite{Fad00}. In Ref.~\cite{FFP99} the quark
contribution to the ``real'' part of the kernel of the generalized BFKL equation, 
${\cal K}_r^{Q \,({\cal R})\,(1)}$ has been calculated for singlet
and octet color representation in the $t$-channel and it has been explicitly
shown that in the octet case the first bootstrap condition~(\ref{boot1})
is fulfilled for arbitrary space-time dimension $D$.

The second bootstrap condition has been considered in the case of
quark impact factors~\cite{FFKP00_Q} and gluon impact factors~\cite{FFKP00_G}.
In the case of quark impact factors, the intermediate states to be considered
in the determination of ${\mathrm Im}{\cal A}_{P^{^{\prime }}P}^{(8,\nu)}$
(see Eq.~(\ref{imp_fact})) are one-quark at 1-loop level and quark-gluon pair 
at Born level. In the case of gluon impact factors, the intermediate states 
to be considered are one-gluon at 1-loop level and two-gluon
and $q \overline q$-pair at Born level. Let us outline the main points of the 
calculation in the simpler case of the quark impact factors. The contribution from
the one-quark intermediate state involves the convolution of the QQR
effective vertex $\Gamma^{c \,(B)}_{QA}$ at Born level with the QQR effective 
vertex at 1-loop level $\Gamma^{c \,(1)}_{QA}$~\cite{FFQ94,FFQ96}
(here A stands for the colliding quark, Q for the quark in the intermediate state).
Differently from $\Gamma^{c \,(B)}_{QA}$, $\Gamma^{c \,(1)}_{QA}$ contains 
also a helicity non-conserving term. The contribution from the quark-gluon 
intermediate state involves the convolution of two $\{QG\}AR$ effective vertices
$\Gamma^c_{\{QG\}A}$ at Born level~\cite{FFQ96}. Again, this contribution to the 
quark impact factor will contain both helicity conserving and non-conserving terms.
In Ref.~\cite{FFKP00_Q} the NLA quark impact factors have been given in terms of 
integrals for arbitrary color state in the $t$-channel and 
non-forward case. For the octet color representation it has been 
explicitly shown that the second bootstrap condition is fulfilled, both for
the helicity conserving and non-conserving parts of the quark impact factors
and for arbitrary space-time dimension $D$.
It must be stressed that the check of the bootstrap for the
helicity conserving part represents a check of correctness of previous
calculations more than a real test for the gluon Reggeization, since this part
is not quite independent of the calculation of the two-loop correction to the gluon 
trajectory. The check of the bootstrap for the helicity non-conserving part is 
instead a really new
check of the gluon Reggeization. The explicit calculation of the integrals appearing
in the expression for the quark impact factors has been performed in the case 
of massless quarks~\cite{FFKP00_Q}.   

The future work includes the check of the first bootstrap condition 
for the part concerning the gluon contribution~\cite{FFP00}. The ingredients 
for this equation are the gluon contribution to the 2-loop Reggeized gluon 
trajectory $\omega_G^{(2)}(t)$~\cite{Fad00}
and the gluon contribution to the ``real'' part of the NLA kernel in the octet
color representation, 
${\cal K}_r^{G \,(8)\,(1)}( \vec q_1,\vec q_2;\vec q)$,
\[
{\cal K}_r^{G \, (8)\,(1)}\left( \vec q_1,\vec q_2;\vec q\right) = 
{\cal K}_{RRG} ^{G \,(8)\,(1)}\left( \vec q_1,\vec q_2;
\vec q\right) + {\cal K}_{RRGG}^{\left( 8\right) \left(B\right) }
\left( \vec q_1,\vec q_2;\vec q\right)\;.
\]
${\cal K}_{RRG} ^{G \,(8)\,(1)}$ depends on the RRG 
effective vertex at Born level $\gamma_{c_1 c_2}^{G (B)}$~\cite{Fad00}
and on the real part of the gluon contribution to the RRG effective vertex 
at 1-loop level $\gamma_{c_1 c_2}^{G\,(G)(1)}$~\cite{Fad00}, known for arbitrary
kinematics only in the $D\rightarrow 4$ limit. The determination of 
$\gamma_{c_1 c_2}^{G\,(G)(1)}$ for arbitrary space-time dimension $D$, in 
terms of uncalculated integrals (that is enough at least for the check of the 
bootstrap) is in progress~\cite{FFP00}.
${\cal K}_{RRGG}^{\left( 8\right) \left(B\right) }$ depends 
on the RRGG effective vertex $\gamma _{c_1c_2}^{GG}$~\cite{Fad00} 
and is known~\cite{FG00,Fad00}.
The final step will be the check that the first bootstrap condition~(\ref{boot1})
is fulfilled also for the gluon part of the kernel of the generalized
non-forward BFKL equation for arbitrary space-time dimension $D$~\cite{FFP00}.
This check is already known to be fulfilled in the physical limit 
$D\rightarrow 4$~\cite{Fad00,FFK00}.

Very recently, another set of ``strong'' bootstrap conditions has been 
derived~\cite{BV00,FFKP00}, based on the assumption of NLA Reggeization
of Reggeon-particle scattering amplitudes. As a consequence of
this assumption, it turns out that the impact factors of any colliding
particle in the octet color representation are proportional to the related PPR
effective vertex, with a universal coefficient function.

\vspace{0.2cm}

\noindent {\large {\bf Acknowledgments}}

The author acknowledges the invaluable pleasure coming from the  
listening of the ``mor\-skoi priboi'' of the Black Sea during the time
of the Conference.

\end{document}